\documentclass{iopconfser}
\usepackage{physics}
\usepackage{mathrsfs}
\usepackage{amsmath}
\usepackage{amssymb}

\begin{document}

\title{On the magnetic 2+1- D space-time and its non-relativistic counterpart}

\author{Sayan Kumar Pal}

\affil{CVV-Institute of Science and Technology,  Chinmaya Vishwa Vidyapeeth (Deemed to be University), Ernakulam, Kerala 686667, India}
%\affil{$^2$Department, Institution, City, Country}

\email{pal.sayan566@gmail.com, sayan.kumar@cvv.ac.in}

\begin{abstract}
	We present here an interesting non-relativistic limit, referred as the Newton-Hooke (NH) limit, of the purely magnetic BTZ solution by starting from the Einstein-Maxwell system in the 2+1 dimensions. The Newton-Hooke limit is different from the Galilean limit in the sense that the former contains an additional parameter $\Lambda$, the cosmological constant over and above the speed of light, c. We show that under this limit, the geodesics of the magnetic BTZ solution reduce to the two-dimensional motion of a charged particle in a normal magnetic field together with the presence of an extra harmonic potential, sometimes called as the Fock-Darwin problem which serves as precursor to model certain condensed matter theories. Our present study has significance in analyzing the symmetries of different dynamical systems, from relativistic and/to non-relativistic theories. Also, we discuss here one of the applications of the generalized (magnetic) $NH_3$ symmetry in the context of Virial theorem, where this symmetry is the symmetry group of the Fock-Darwin problem mentioned above.

\end{abstract}

\section{Introduction}

It is now well-known that the geodesic equation of motion of the free neutral particle in $AdS_3$ geometry takes the form of a simple oscillatory system under a certain nonrelativistic limit, called as the Newton-Hooke (NH) limit \cite{gibbons,gibb,ellis,aldrovandi}. This limit is defined as the nonrelativistic weak ($v<<c$) limit of Einstein gravity together with the vanishing limit of the negative cosmological constant of an otherwise $AdS$ spacetime geometry in the following manner:  $c \rightarrow \infty$ and $\Lambda \rightarrow 0$, precisely such that $c^2\Lambda \rightarrow -\omega^2$ is held finite. This is known as the Newton-Hooke limit of the relativistic theory. Furthermore, from the group theoretical perspective, the relativistic $AdS_3$ can be mapped to this nonrelativistic $NH_3$ group through Inonu-Wigner group contraction scheme, see \cite{epjc} and references therein. The $AdS_3$ metric has the following form in the global coordinates,
\begin{equation}\label{firsteqnads}
	ds^2=-(1 - \Lambda r^2) c^2dt^2 + \frac{dr^2}{(1-\Lambda r^2)} +r^2 d\theta^2
\end{equation}
The equation of motion in the precise NH limit however reduces to the standard equation of a two-dimensional oscillator,
\begin{equation}
	\ddot{\textbf{r}}=-\omega^2 \textbf{r} ~~ ; ~~~\text{where}~~ \omega^2=-c^2\Lambda
\end{equation}
for, $~\Lambda =- \frac{1}{R^2}$ being negative and, $R$ is the AdS radius. Alternatively, the energy function can be shown to be reducing to,
\begin{equation}
	\epsilon= \frac{p^2}{2m} + \frac{1}{2}m\omega^2 r^2
\end{equation}
 This $NH_3$ geometry has various applications in the context of astrophysical systems and cosmology which have been discussed in \cite{gibb, ellis, freed, vir}. 
On the other hand, recently we have found an interesting generalization of this $NH_3$ symmetry algebra, which we refer as the generalized $-NH_3$ Lie algebra \cite{epjc} (see Appendix). The appearance of this ``generalized NH algebra" was noticed while studying the nonrelativistic symmetry structure of the exotic oscillator model \cite{sir2,stichel} -
\begin{equation}\label{eqn1}
	\ddot{x}_i-m\omega^2\theta \epsilon_{ij}\dot{x}_j + \omega^2 x_i =0 ~~,
\end{equation}
a system that manifestly exhibits a non-standard deformation in the symplectic structure. Such deformed oscillators are generally thought to represent a low-energy nonrelativistic effective system that is believed to be originally derived from physics near Planck scales \cite{sirrb, stichel}. Having said that, for 2D systems which is the case of our present discussion, the deformation parameter $\theta$ may be ascribed to the anyonic spin parameter of particles in 2D under certain considerations \cite{nair, duval01, stichel}. Again more physically and significantly, from the symplectic viewpoint, deformations in the symplectic structure naturally arise inside solids when electron dynamics is described in terms of Wannier coordinates and Bloch wavefunctions \cite{xiao}. Furthermore, an interesting thing about the exotic oscillator system is that it can be mapped to the classic Landau motion problem with an additional 2D harmonic oscillator potential, which sometimes is referred as the Fock-Darwin problem in the literature \cite {dar, fock, peres} or also referred as the non-degenerate Landau problem, if we identify the deformation (exotic) parameter as, 	$\theta=\frac{qB}{m^2\omega^2}$ \cite{epjc}. Here, $B$ is the normal magnetic field to the confining 2D motion of the charged particles in the classic Landau problem.

Now as a matter of fact, the above-mentioned generalized $NH_3$ algebra could not be shown to be descending from the standard $AdS_3$ group through a group contraction \cite{epjc}. This drives us to look for a particular extension of the pure $AdS_3$ spacetime that might reduce to the above-mentioned generalized $NH_3$ nonrelativistic geometry and the corresponding Lie-algebra. Being motivated by the above facts, in the present work we shall be studying the magnetic BTZ spacetime which is a generalization of the BTZ solution found by several people \cite{clement0, welch, dias, cataldo} as a purely magnetically charged static space-time solution in 2+1-D and then show that the Newton-Hooke limit of this relativistic theory in 2+1-dimensions yields us nonrelativistic dynamics that exhibits the generalized Newton-Hooke symmetries mentioned before. This is one of the key results of our present article.\\

Over the last two decades or so, people have found renewed interests in cosmology with nonrelativistic physics \cite{gibbons, gibb, ellis, peebles}. This is attributed to quite a few associated intriguing developments over the years such as understanding the FLRW and Raychaudhuri's equations from Newton-Hooke (NH) physics \cite{ellis}, and the black hole (BH) thermodynamics as fluid relations \cite{paddy}, etc. This observation serves as another important motivation for the present work apart from the theoretical interests mentioned above. Thus in this article, we will also analyse a possible implication of a Newton-Hooke-like nonrelativistic space-time and algebra posited at the fundamental level in the context of astrophysical systems in the spirit of \cite{gibb, vir}. Actually, in our present work, we are curious to find out the encompassing relativistic space-time geometry of dynamical systems that manifests the previously-mentioned nonrelativistic generalized $NH_3$ symmetry and this process might lead to uncovering newer features of the relativistic system, see \cite{jcap0, jcap}. Thus, the paper is arranged as follows: section 2 provides a review of the derivation of the magnetic BTZ metric, section 3 considers particle dynamics in the relativistic magnetic BTZ space-time and then we carry out the Newton-Hooke limit of the theory, in section 4 assuming a background generalized NH space-time exhibiting the generalized NH symmetries, we analyse its implications in the context of the Virial theorem, and, finally we conclude in section 5.

\section{Magnetic and non-rotating BTZ metric}
In this section, since we are primarily interested in probing the nonrelativistic limit, we need to review the derivation of the magnetic BTZ space-time solution found previously by others \cite{clement0, welch, dias, cataldo} by retaining all quantities with their actual dimensions and keeping $c$ explicit throughout, which are generally not considered in most of the literature. To that end, we begin with the Einstein-Hilbert action for the following case of an Einstein-Maxwell system in 2+1-dimensions:
\begin{equation} S_{E-H}=\frac{1}{4c}\int d^3x \sqrt{|g|} \bigg(\frac{R-2\Lambda}{k}-\frac{1}{\mu_0}F^{\mu\nu}F_{\mu\nu}\bigg)
\end{equation}
Extremizing the action yields Einstein's equation in the presence of the cosmological constant $\Lambda$ and the electromagnetic field $F_{\mu\nu}$ as the matter contribution,
\begin{equation}\label{Einseqn}
	R_{\mu\nu}-\frac{1}{2}Rg_{\mu\nu} + \Lambda g_{\mu\nu}=k T_{\mu\nu} ~~;~ k=\frac{4\pi G_3}{c^4}
\end{equation}
where the components of the energy-momentum tensor $T_{\mu\nu}$
are given by,
\begin{equation}\label{Energymom}
	T_{\mu\nu}=\frac{1}{\mu_0}\bigg(-\frac{1}{4}g_{\mu\nu}F^{\rho\sigma}F_{\rho\sigma}+g^{\rho\sigma} F_{\mu\rho}F_{\nu\sigma}\bigg)
\end{equation} 
Here, note that we are following the $(-,+,+)$ convention for the metric as well as the popular (+,+,+) convention of MTW \cite{mtw}.
%$Furthermore, one also has Maxwell's equation in the absence of a source as,
%\begin{equation}\label{maxwelleqn}
%\nabla_\nu F^{\mu\nu}=0
%\end{equation}
Note that Newton's constant $G_3$ in (2+1)-D has the dimension of $\frac{L^2}{MT^2}$ and absolute permeability $\mu_0$ possesses $\frac{M}{A^2T^2}$ (here $A$ denotes ampere), which has been fixed using the standard dimension of the electric field together with the dimension of the action functional. Now, contraction with $g^{\mu\nu}$ in (\ref{Einseqn}) leads to,
$R=6\Lambda-2kT=6\Lambda$ (for a traceless energy-momentum tensor). Therefore, this simplifies (\ref{Einseqn}) to,
\begin{equation}\label{Eins2.}
	R_{\mu\nu}=k T_{\mu\nu}+2 \Lambda g_{\mu\nu}
\end{equation}
Next, consider a general form of a metric for the coordinates ($t, r$, and $\theta$) in (2+1) dimensions :
\begin{equation}\label{metric1}
	ds^2=-a(r)c^2dt^2 + b(r)dr^2+h^2(r)d\theta^2
\end{equation}
where $a(r)$, $b(r)$ are dimensionless functions of the radial coordinate $r$, while $h^2(r)$ is also a function of $r$ albeit with $L^2$ dimension. As our present interest lies in obtaining the static (i.e. non-rotating) solutions in the BTZ spacetime and purely in the magnetic field solution, we consider the electromagnetic field tensor $F_{\mu\nu}=\partial_\mu A_\nu-\partial_\nu A_\mu$ in the following simpler form,
\begin{equation}F_{\mu\nu}=
	\begin{pmatrix}
		0 & 0 & 0\\
		0 &  0 &  -rB\\
		0 & rB & 0
	\end{pmatrix} \label{emtensor}
\end{equation}
We now provide an outline of finding the unknown metric coefficients $a(r), b(r), h^2(r)$ by consistently solving the Einstein-Maxwell system of equations (\ref{Einseqn}) and (\ref{maxwe}). The different components of the Ricci tensor $R_{\mu\nu}$ and the energy-momentum tensor $T_{\mu\nu}$ are evaluated from which we have explicitly,
\begin{equation}\label{Trelation}
	\frac{T_{11}}{b}-\frac{T_{22}}{h^2}=0
\end{equation}
Using (\ref{Eins2.}) in the above equation, we get,
\begin{eqnarray}\label{Rrelation}
	\frac{R_{11}}{b}-\frac{R_{22}}{h^2}&=&0  \nonumber \\
	or, ~ \frac{a'}{a}+\frac{b'}{b}+2\frac{h'}{h} &=& 2\frac{a''}{a'}  \label{1steqn}
\end{eqnarray}
Integrating the above, we obtain -
\begin{eqnarray}
	abh^2&=&\lambda_1 a'^2
\end{eqnarray}
where, $\lambda_1$ is a constant of integration independent of $r$.\\

Next, we consider Maxwell's equations,
\begin{equation}\label{maxwe}
	\partial_\nu F^{\mu\nu}+\Gamma^{\nu}_{\nu\rho}F^{\mu\rho}=0
\end{equation}
Putting $\mu=2$ and simplifying to arrive at,
\begin{eqnarray}
	\partial_r F^{21}+\frac{a''}{a'}F^{21}=0 \nonumber
	\implies  F^{21}&=&\frac{\lambda_2}{a'} \nonumber \\
	\text{Thus}, ~B=\frac{1}{r}F_{21}=\frac{bh^2}{r} \frac{\lambda_2}{a'} \label{magneticfield}
\end{eqnarray}
where $\lambda_2$ is another constant of integration. Also, we have $R_{00}=k T_{00}+2 \Lambda g_{00}$ (\ref{Eins2.}).  Explicitly writing this to obtain,
\begin{equation}\label{einst00}
	- \frac{a'}{2}(\frac{a'}{a}+\frac{b'}{b})+\frac{a'h'}{h}+a''=-\frac{k}{\mu_0}\frac{ar^2}{h^2}B^2-4\Lambda ab 
\end{equation}
Adding (\ref{1steqn}) and (\ref{einst00}), we get upon simplifying and integrating,
\begin{eqnarray}
	%2hh'&=&\frac{2 k\Lambda \lambda_1^2 \lambda_2^2}{\mu_0}\frac{r}{1-\Lambda r^2}+8\Lambda^2 \lambda_1 r \nonumber \\
	h^2&=&4\Lambda^2\lambda_1r^2+\frac{k\lambda_1^2 \lambda_2^2}{\mu_0}\log{|1-\Lambda r^2|}+\log{C_1}
\end{eqnarray}
where we have used, $a= 1- \Lambda r^2 $ from asymptotic conditions (\ref{firsteqnads}). Now, in the absence of a magnetic field $B$, the spacetime becomes pure $AdS_3$. In other words, if $\lambda_2=0$, $h^2(r)$ must go to $r^2$. This fixes the constant of integration $C_1=1$ and $\lambda_1=\frac{1}{4\Lambda^2}$. The only undetermined constant at this stage is $\lambda_2$ which is obviously related to the magnetic field strength $B$ and will be discussed in sub-section 3.1.
Therefore, (\ref{gphi1}) turns into,
\begin{eqnarray}\label{gphi1}
	h^2(r)&=&r^2+\frac{k \lambda_2^2}{16\mu_0 \Lambda^4}\log{|1-\Lambda r^2|}
\end{eqnarray}
Therefore, the different metric components are finally determined to be -
\begin{equation}\label{metriccompo1}
	a(r)=1-\Lambda r^2 ~;~~ h^2(r)= r^2 + \chi^2 \log{|1-\Lambda r^2|} ~;~~ \text{and},~ b(r)=\frac{r^2}{ah^2}
\end{equation}
where, 
\begin{equation}\label{magstrength}
	\chi=\sqrt{\frac{k}{\mu_0}}\frac{\lambda_2}{4 \Lambda^2}=\frac{2\lambda_1\lambda_2}{c^2\eta}=\frac{\lambda_2}{2 \Lambda^2 c^2 \eta}~~~;~\eta=\sqrt{\frac{\mu_0}{\pi G_3}}
\end{equation}
Basically, $\eta\chi$ denotes the strength of the magnetic field in this spacetime, as we will shortly find out. Hence, the metric (\ref{metric1}) of the magnetically charged BTZ spacetime takes the form:
\begin{equation}\label{metricfinal}
	ds^2=-(1 - \Lambda r^2) c^2dt^2 + \frac{r^2~dr^2}{(1-\Lambda r^2)(r^2 + \chi^2 \log|1-\Lambda r^2|)} + (r^2 + \chi^2 \log|1-\Lambda r^2|)d\theta^2
\end{equation}
This metric has appeared previously in a number of works \cite{welch, dias, cataldo}. It is obvious from the above expression that $\chi$ has the dimension of $L$. Indeed, this is true which may be checked explicitly by directly substituting the dimensions of the various quantities.
 
Now that we have obtained the form of the metric of a purely magnetic BTZ spacetime with all quantities having the proper dimensions, we can now proceed to analyse the particle dynamics. However, before we investigate particle motions, let us take a look into the magnetic field obtained.

%%%%%%%%%%%%%%%%%%%%%%%%%%%%%%%%%%%%%%%%%%%%%%%%%%%%%%%%%%%%%%%%%%%%%%%%%%%%%%%%%%%%%%%%%%%%%%%%%%%%%%%%%%%%%%%%%%%%%%%%%%%%%%%%%%%%%%%%%%%%%%%%%%%%%%%%%%%%%%%%%%%%%%%%%%%%%%%%%%%%%%%%%%%
\section{Particle dynamics in the magnetic BTZ space-time}
We will now want to analyse the dynamics of a relativistic charged particle moving in the magnetic BTZ space-time. To that end, we write down the Lagrangian $L$ of the relativistic charged particle moving in the background magnetically charged BTZ spacetime (\ref{metricfinal}) -
\begin{equation}
	L=\frac{1}{2}g_{\mu\nu} \dot{x^{\mu}} \dot{x^{\nu}} + qA_{\mu}\dot{x^{\mu}}
\end{equation}
\begin{equation}
	\implies L=\frac{1}{2}\bigg(-a(r)c^2\dot{t}^2+b(r)\dot{r}^2+h^2(r)\dot{\theta}^2\bigg)+qA_{\theta}\dot{\theta}
\end{equation}
Specifically, the geodesic equation for the r-coordinate is computed to be -
\begin{equation}
	\ddot{r} = \frac{1}{2}\bigg(-\frac{a'
	}{b}c^2\dot{t}^2-(\frac{2}{r}-\frac{a'}{a}-2\frac{h'}{h})\dot{r}^2\bigg)+ \frac{hh'
	}{b}\dot{\theta}^2-q\frac{c^2\eta\chi\Lambda}{ab}r\dot{\theta}
\end{equation}
Using the canonical quantities, energy, $E=-ac^2\dot{t}$ and modified angular momentum $\tilde{l}=l+qA_{\theta}$ ($l=h^2\dot{\theta}= ~$`mechanical' angular momentum), one can rewrite -
\begin{equation}\label{reom1}
	\ddot{r} = -\frac{a'
	}{2a}\frac{1}
	{ab}\frac{E^2}{c^2}-(\frac{1}{r}-\frac{a'}{2a}-\frac{h'}{h})\dot{r}^2+ \frac{h'
	}{h}\frac{a}{r^2}(\tilde{l}-qA_{\theta})^2-q\frac{c^2\eta\chi\Lambda }{r}(\tilde{l}-qA_{\theta})  
\end{equation}
where, the functions $a(r), b(r)$, $h(r)$ and $A_{\theta}$ are as given previously, whereas quantities like $c, E, \eta, \chi$, etc are constants.\\

Now quite anticipatedly, the equation of motion looks slightly complicated and we do not get into the analysis of this equation of motion. Rather, as mentioned at the outset, we want to confine ourselves to investigate about the nonrelativistic limit of this theory. Let us go through the same now. 
\subsection{Nonrelativitic limit}
For the sake of convenience, we further rewrite (\ref{reom1}) as,
\begin{equation}\label{reom2}
	\ddot{r} = \frac{\Lambda h^2}
	{ra}\frac{E^2}{c^2}-(\frac{1}{r}+\frac{\Lambda r}{a}+\frac{\Lambda\chi^2 r}{ah^2}-\frac{r}{h^2})\dot{r}^2+\frac{1}{rh^2}(a-\Lambda\chi^2)(\tilde{l}-qA_{\theta})^2-q\frac{c^2\eta\chi\Lambda }{r}(\tilde{l}-qA_{\theta})  
\end{equation}
Next, we perform the relativistic expansion of energy (i.e. $E=\epsilon + mc^2$, where $\epsilon$ is the nonrelativistic energy function) and notice the following fate of the metric components $a(r)$ and $h^2(r)$ under the previously mentioned NH limit ( i.e. $c \rightarrow \infty$ and $\Lambda \rightarrow 0$, precisely such that $c^2\Lambda \rightarrow -\omega^2$ is finite):
\begin{equation}
	a(r) \rightarrow 1 ~;~ ~ c^2 a(r) \rightarrow c^2+\omega^2 r^2 ~~ \text{and},~ h^2(r)\rightarrow r^2
\end{equation}
Accordingly, taking the NH limit in all the terms of the final cast form of the equation of motion (\ref{reom2}), we arrive at -
\begin{equation}\label{HOandB}
	\ddot{r} = -\omega^2 r + \frac{\tilde{l}^2}{r^3}-\frac{q^2B^2}{4}r
\end{equation}
where now, $\tilde{l}=l+q\frac{B}{2}r^2$ (and $l=r^2\dot{\theta}$).

Interestingly, the above is the equation of motion of a nonrelativistic charged particle confined to a plane under the influence of a normal magnetic field and an additional 2D harmonic potential (i.e. the Fock-Darwin problem, also regarded as the nondegenerate Landau problem) when expressed in terms of the plane-polar coordinates $r, \theta$ rather than the more familiar form with the cartesian coordinates (see Appendix).

Also, the trajectory equation is given by, after a straightforward calculation,
\begin{equation}
	\bigg(\frac{dr}{d\theta}\bigg)^2=-a\frac{h^4}{r^2}+\frac{c^2 a}{r^2}\frac{h^6}{\tilde{l}^2}+\frac{h^6}{r^2\tilde{l}^2}\frac{E^2}{c^2}
\end{equation}
A similar calculation of the NH limit leads to the following equation of the trajectory of the corresponding NR system:
\begin{equation}\label{BTZm}
	\bigg(\frac{dr}{d\theta}\bigg)^2=-r^2-\frac{m^2\omega^2r^6}{\tilde{l}^2}+\frac{2m\epsilon r^4}{\tilde{l}^2}
\end{equation}
Again, this reiterates the fact that the theory of the NH limit of magnetic BTZ spacetime is equivalent to the system of a nonrelativistic charged particle confined to a plane under the influence of a normal magnetic field and an additional 2D harmonic potential. Finally, we want to elaborate on the nature of the obtained magnetic field as there seems to be different opinions in the literature \cite{clement0, welch, dias, cataldo}.

\subsection{Nature of the magnetic field}
Just in order to differentiate between the magnetic field obtained from solving the Einstein-Maxwell system and the analogous nonrelativistic magnetic field, we ascribe $B_{AdS}$ for the magnetic field in (\ref{magneticfield}) or (\ref{emtensor}) and $B_{Landau}$ for the analogous magnetic field of the popular nonrelativistic Landau motion. 
The gauge potential 1-form has the following form:
\begin{equation}
	A=\frac{c^2\eta\chi}{2}\log|1-\Lambda r^2|d\theta \nonumber \\
	\implies A_{\theta}=\lambda_1\lambda_2\log|1-\Lambda r^2|\label{Atheta}
\end{equation}
Thus, the magnetic field is:
\begin{eqnarray}
	B_{AdS}=\vec{\nabla} \cross \vec{A}=\frac{1}{r}\partial_r A_{\theta}=\frac{bh^2}{r} \frac{\lambda_2}{a'} \nonumber \\
	\implies B_{AdS}=\frac{r\lambda_2}{aa'}=-\frac{\lambda_2}{2\Lambda}\frac{1}{1-\Lambda r^2}\label{Atheta}
\end{eqnarray}
Choosing $\lambda_2=-2\Lambda B_{Landau}$ where, $B_{Landau}$ denotes some constant magnetic field (the same field that appears in the classic Landau problem and hence the name). Hence, we finally obtain the magnetic field in $AdS_3$ space to be given by:
\begin{equation}\label{brelate}
	B_{AdS}=\frac{B_{Landau}}{1-\Lambda r^2}
\end{equation}
And thus, in the NH-limit, $B_{AdS}$ reduces to the magnetic field in the Landau system, $B_{Landau}$.

\section{Application:} 
In this pen-ultimate section, we would just want to mention an implication of the generalized $NH_3$ structure on the Virial Theorem. Here, let us not bother about the specific physical realization of this symmetry, but rather we take it as a fundamental symmetry of certain systems and try to see its effect on the Virial Theorem. In Newtonian physics, 
\begin{equation}
	a_i=\frac{d^2{x_i}}{d{t}^2}=-\frac{\partial{\phi_{net}}}{\partial{x^i}}
\end{equation}
For the generalized NH spacetime, the net potential turns out to be given as $\phi_{net}= \frac{\omega^2}{2}x_j^2 - \frac{qB}{m}\epsilon_{jk}\dot{x}_jx_k$.\\
Let us now consider a density function $f=f(x_i,v_i,t)$.
The total probability conservation (Liouville's theorem) readily gives us the following equation:
\begin{equation}
		\frac{df}{dt}=\frac{\partial{f}}{\partial{t}}+ v^i\frac{\partial{f}}{\partial{x^i}}+\frac{dv_i}{dt}\frac{\partial{f}}{\partial{v^i}}=0
\end{equation}
	\begin{equation}
		\implies \frac{\partial{f}}{\partial{t}}+ v^i\frac{\partial{f}}{\partial{x^i}}+\frac{\partial{f}}{\partial{v^i}}\bigg(-\partial^i\phi_N+\frac{\Lambda}{3} x^i - \frac{qB}{m}\epsilon^{ij}\dot{x}^j \bigg)=0
\end{equation}
where, $\phi_N$ is the Newtonian gravitational potential. 
 From here, one can derive a modified Virial relation based on the generalized NH structure given as,
\begin{equation}
	\frac{1}{2}\frac{\partial ^2 I_{ij}}{\partial t^2}= 2 K_{ij} + W_{ij} + \frac{\Lambda}{3}I_{ij}-\frac{qB}{2m}\epsilon_{il}\frac{\partial I_{jl}}{\partial t}
\end{equation}
where, 	$I_{jk}$ is the moment of inertia tensor, $K_{jk}$ is the kinetic energy tensor and $W_{jk}$ is the rotation tensor and are respectively given by,
\begin{equation}
I_{jk}=\int \rho x_j x_k d^3x ~~;~\rho =\int fd^3v
	\end{equation}
\begin{equation}
	K_{jk}=\frac{1}{2}\int \rho \overline{v_jv_k} d^3x
\end{equation}
\begin{equation}
	W_{jk}=-\int \rho x_k \frac{\partial{\phi_N}}{\partial{x^j}} d^3x
\end{equation}
The first three terms have been already discussed in \cite{gibb, ellis}, whereas the last term is worth-noticing and represents the contribution due to the magnetic field. For sake of conciseness, we have omitted here the detailed derivation of the modified Virial relation which itself is bit involved and shall be presented in a future work. Also, this new symmetry can have applications in nonrelativistic cosmology \cite{ellis, freed} for example in the nonrelativistic motion of a group of N-point particles in an expanding homogeneous isotropic universe with the cosmic time and obtain a modified Raychaudhuri equation and correspondingly get the FLRW equation.

\section{Conclusion}
In this work, we have explored the Newton–Hooke (NH) limit of the purely magnetic BTZ spacetime in 2+1 dimensions, starting from the Einstein–Maxwell system. We demonstrated that, under this limit, the relativistic geodesic equations reduce to the effective dynamics of a charged particle moving in a plane under a uniform magnetic field supplemented by a two-dimensional harmonic potential — the well-known Fock–Darwin system, also sometimes referred as the non-degenerate Landau problem. This identification establishes a concrete connection between a relativistic gravitational background and a nonrelativistic mechanical system (a condensed matter analogue), thereby enriching the scope of NH-type limits beyond the Galilean framework. Additionally, we believe that our analysis provides clarification on the physical nature of the magnetic field in the magnetically charged BTZ solution through our provided relation in (\ref{brelate}).

Our analysis also highlighted the role of the generalized $NH_3$ symmetry, which naturally arises in this context and governs the dynamics of the resulting system. In particular, we discussed its implications for the Virial theorem, where the additional magnetic contributions appear explicitly, suggesting further applications in both nonrelativistic cosmology and  effective descriptions of planar charged systems.

One of the shortcomings of the present work is the unavailability of the mapping between the two theories - relativistic and nonrelativistic at the level of the symmetry group generators since it is quite challenging to analytically compute the time-dependent symmetry generators of the purely magnetic BTZ system because of the complicated form of the geodesic equation in (\ref{reom2}). However, work is in progress along these lines and we hope to report the results in the near future.

The present results open several directions for future research. One could study the entropic and entanglement aspects of such space-time solutions in 2+1-D, and try to correlate it to the entropy of the obtained nonrelativistic system \cite{skp}. Secondly, as mentioned just before, we are in the process of analysing the possibility of the existence of new symmetry generators for the magnetic BTZ solution in 2+1-dimensional Einstein gravity.

Thus, to sum up, a geometrical interpretation of the generalized $NH_3$ symmetry, as the NH limit of magnetic BTZ space-time solution has been now provided. Overall, our findings illustrate how nonrelativistic limits can reveal hidden structures in gravitational theories and connect them to well-studied dynamical models in lower dimensions, thereby strengthening the bridge between relativistic geometry, nonrelativistic dynamics, and condensed matter analogues. \\

\textbf{Acknowledgements:} The author expresses gratitude to the organizers of the conference ISQS29 held in Czech Technical University in Prague, where this work was presented. Also, the author is grateful to the Indian Statistical Institute (ISI) for a visiting scientist position when this work was started. Sincere thanks are due to Subir Ghosh for valuable discussions during the various stages of the work. This work is dedicated to the loving memory of Prof. A.P. Balachandran.

\section*{Appendix}
For the sake of completeness, we briefly present the generalized $\overline{\mathcal{NH}}_3$ algebra.
The equation for the exotic oscillator is :
\begin{equation}\label{eqn1}
	\ddot{x}_i-m\omega^2\theta \epsilon_{ij}\dot{x}_j + \omega^2 x_i =0
\end{equation}
Now as mentioned before, the above system can be mapped through a canonical transformation (or an unitary transformation at the quantum level) to the non-degenerate Landau problem (or the Fock-Darwin problem) described by the following Lagrangian:
\begin{equation}{\label{lagrangian1}}
	L=\frac{1}{2}m\dot{x_i}^2- \frac{1}{2}m\omega^2 x_i^2 +\frac{qB}{2}\epsilon_{ij}x_i \dot{x}_j
\end{equation}
on identifying,
\begin{equation}\label{map}
	\theta=\frac{qB}{m^2\omega^2}
\end{equation}
The gauge potential chosen here is the symmetric gauge, $A_j=\frac{B}{2}\epsilon_{ij}x_i$. Here, $B$ denotes the constant normal magnetic field. The equation of motion is :
\begin{equation}\label{eqnlandau}
	\ddot{x}_i- \frac{qB}{m} \epsilon_{ij}\dot{x}_j + \omega^2 x_i =0
\end{equation}
which exactly resembles (\ref{eqn1}). And, the equation of motion in the plane polar coordinates $r, \theta$ will take the form discussed in (\ref{HOandB}) in section 3.1.\\

The symmetry algebra of the above kind of systems, which we have regarded as the generalized Newton-Hooke algebra, $\overline{\mathcal{NH}}_3$ takes the form:

\begin{equation}
	\{K_i,K_j\}=\theta m^2 \epsilon_{ij} \nonumber
\end{equation}
\begin{equation}
	\{T_i,T_j\}=\theta m^2 \Omega^2 \epsilon_{ij} \nonumber
\end{equation}
\begin{equation} \label{newalgebra}
	\{K_i,T_j\}=(1+\frac{ m^2 \omega^2 \theta^2}{2})m\delta_{ij} \nonumber
\end{equation}
\begin{equation}\label{NCPoisson}
	\{H,K_i\}=-T_i + \beta \epsilon_{ij}K_j ~ ~;~\{H,T_i\}=\Omega^2 K_i+\beta \epsilon_{ij}T_j 
\end{equation}
Note that the spatial translations do not commute here unlike in the case of the ordinary $NH_3$ algebra (symmetry group of the standard 2-D harmonic oscillator system) and boosts also don't commute in this case. This generalized (or magnetic) $NH_3$ Lie-algebra is reminiscent of a second centrally-extended Newton-Hooke algebra $\tilde{\mathcal{NH}}_3$ but not identical and is a larger symmetry algebra/group than the standard Newton-Hooke (for details, refer \cite{epjc}). Here, the deformation parameter $\theta$ may represent the anyonic spin in two dimensions which can be any real number \cite{stichel, nair}. The radius of curvature of the space $(R=\frac{1}{\sqrt{-\Lambda}})$ is represented by the frequency $\omega$ of the NC oscillator and together with the NC parameter $\theta$ is responsible for the noncommuting translations which possibly indicate to a presence of curvature in the space of motions.\\

%\section*{References}
%\bibliographystyle{iopart-num}
%\bibliography{references}

\end{document}